\documentstyle[aps,preprint]{revtex}
\begin{document}
\draft

\preprint{KIAS-P97012, SNUTP-97-163}

\title{Infinite Lorentz boost along the M-theory circle and
non-asymptotically flat solutions in supergravities}
\author{Seungjoon Hyun $^{(a)}$, Youngjai Kiem $^{(b)}$, 
       and Hyeonjoon Shin $^{(c)}$}
\address{ $^{(a)}$ Department of Physics, Kyunghee University, 
                 Seoul 130-701, Korea\\
          $^{(b)}$ School of Physics, KIAS, Seoul 130-012, Korea\\
          $^{(c)}$ Department of Physics Education, 
                 Seoul National University, Seoul 151-742, Korea \\
              (  sjhyun@nms.kyunghee.ac.kr, ykiem@chep6.kaist.ac.kr,
                hshin@phya.snu.ac.kr )}
\maketitle

\begin{abstract}
Certain non-asymptotically flat but supersymmetric classical 
solution of the type IIA supergravity can be interpreted as 
the infinitely-boosted version of the D-particle solution 
along the M-theory circle.  By a chain of $T$-dual 
transformations, this analysis also applies to yield  
non-asymptotically flat solutions from the asymptotically flat 
and (non)-extremal solutions with intersecting D-strings and 
D five-branes of the type IIB supergravity compactified on a 
five-torus.  Under $S$-duality, the non-asymptotically flat 
solutions in this context can in particular be used to describe 
the (1+1)-dimensional CGHS type black holes via spontaneous 
compactifications.
\end{abstract}


\newpage

\section{Introduction}

D-branes and string dualities revolutionized our understanding of 
the non-perturbative aspects of the string theory and black hole
physics \cite{polchinski}. 
Black holes in various types of supergravities, which
are in turn the low energy limit of the string theories, are 
connected to each other via string dualities, 
and in some cases microscopic descriptions in terms 
of D-branes are possible \cite{cvetic}.
The emergence of M-theory, whose low energy limit is believed
to be described by the eleven dimensional supergravity, considerably
enriches these recent developments.  
The most notable current approach to realize the quantum
version of M-theory is the Matrix theory \cite{bfss}.  In its 
formulation, choosing the infinite momentum frame along the 
extra M-theory circle is a key technical tool.  

In the long distance limit, D-branes are described by the
black hole type brane solutions of supergravities.  In obtaining
these solutions, it is conventionally required that the space-time
is asymptotically flat.  
In \cite{hyun}, however, it has been shown that by
taking the light-cone compactification for each of these D-brane
solutions, it is possible to find non-asymptotically flat 
and supersymmetric solutions which result also from
appropriately setting a number of constants of 
integration\footnote{This kind of phenomena has been observed in
a slightly different context in \cite{kpl} as well.}.  Some of 
these non-asymptotically flat solutions, via $U$-duality, give the
geometry of lower dimensional, in particular two and three
dimensional, black holes \cite{hyun}.  The appearance of the 
lower dimensional black holes tensored with a torus and/or sphere
has been observed in a different context.  For the NS five-branes,
it was noted in Ref. \cite{malda} that the metric near the 
horizon approaches the two dimensional black hole of 
Callan-Giddings-Harvey-Strominger (CGHS) \cite{cghs}.

The relationship between the D-brane solutions and the
non-asymptotically flat solutions is the main focus of this paper;
in the context of the eleven dimensional supergravities, we find
that they are related to each other by an infinite Lorentz boost
along the M-theory circle.  We illustrate this property in 
section II for the simplest case of D-particles.  Our results
in section II are not new in a sense that
the infinite Lorentz boost along the M-theory circle has been
recently investigated in \cite{seiberg}, and our presentation
adopts the results of that reference.  The novelty here is our
explanation that the difficulties usually associated with 
non-asymptotically flat solutions can be resolved in the
context of the eleven dimensional supergravity and under the
infinite Lorentz boost.  Via a chain of $T$-dualities,
the consideration in section II can be extended to arbitrary
D $p$-branes.  In section III, we consider the application of the
infinite boost for the D 5-branes in the case where we have
intersecting D-strings and D 5-branes with a Kaluza-Klein momentum
along the circle where D-strings are wrapped.  For each
asymptotically flat and (non)-extremal solution, we show that
we can obtain a non-asymptotically flat solution which has
the same space-time structure near the horizon as the
asymptotically flat (non)-extremal solutions.  Furthermore, we find that
these non-asymptotically flat, 
(non)-extremal solutions of the type IIB supergravity
turn into lower dimensional black holes, such as
the CGHS black holes, under $S$-duality 
via a spontaneous compactification.  When combined with the
Lorentz transformation along the M-theory circle, the $U$-dual
multiplets of the black holes in supergravities get enlarged 
to include black holes in differing space-time dimensions.

\section{Preliminary: Infinitely-boosted D-particle solutions}

The D-particle solution of the ten-dimensional type IIA supergravity
can be written as follows.  We have the ten-dimensional
metric 
\begin{equation}
ds^2_{10} = - \frac{1}{\sqrt{f}} dt^2 + \sqrt{f} 
    (dx_1^2 + \  \cdots \  + dx_9^2  ),
\label{d0}
\end{equation}
and the dilaton $\phi$ and the R-R one-form gauge field
$A$ are given by
\[ e^{2 \phi} = f^{3/2} \ \ , \ \
   A_t = 1 - \frac{1}{f} \]
where 
\[ f = 1 + \frac{Q}{R_s^2} \frac{1}{r^7} .\]
Here $f$ represents a nine-dimensional harmonic function
and $r^2 = x_1^2 + \  \cdots \ + x_9^2$.  
The integer $Q$ counts the number of D-particles, and
we are using a unit where the eleven dimensional Einstein
action has the unit coefficient in front of the curvature
tensor.  The ten dimensional gravitational constant
is therefore inversely proportional to the eleventh circle radius
$R_s$, and the energy of one D-particle is given by
$E = 1/ (gl_s ) = 1/ R_s$, where $g$ is the string coupling
constant and $l_s$ is the ten dimensional string scale.

The inclusion of the first term, 1, in the harmonic function
$f$ is necessary if we require that the space-time is
asymptotically flat.  On the other hand, under a different 
choice of the constants  of integration, we have solutions 
of the form
\begin{equation}
ds^2_{10} = - \frac{1}{\sqrt{h_0}} dt^2 + \sqrt{h_0} 
    (dx_1^2 + \ \cdots \ + dx_9^2 ) 
\label{non-flat}
\end{equation}
along with
\[ e^{2 \phi} = h_0^{3/2} \ \ , \ \ 
   A_t =  - \frac{1}{h_0} \]
where $h_0$ is another harmonic function given by
\[ h_0 = \frac{k}{r^7} . \]
This solution, although it satisfies the ten-dimensional
equations of motion and supersymmetric, is clearly 
non-asymptotically flat.  There are at least two serious
problems that prevent us from regarding the solution
(\ref{non-flat}) as a physically sensible one.  First, 
the limit $k \rightarrow 0$ is singular.  In contrast,
in the case of the D-particle solution (\ref{d0}), the limit
$Q \rightarrow 0$ simply represents the flat vacuum
solution.  Secondly, and more seriously, the ADM type mass
of the solution (\ref{non-flat}) is difficult to define
due to the non-asymptotic flatness 
in the long distance limit, unlike the D-particle solution.
In fact, the second problem is related to the first problem, since
ideally we hope to compute the energy of non-zero $k$ solution
relative to the $k= 0$ solution.

Even with these difficulties, the solutions (\ref{non-flat})
are interesting; for the dynamical processes in the background 
geometry  of (\ref{non-flat}), the ADM  type mass might be 
defined to give a finite answer for the mass, as is familiar 
from the lower dimensional gravity theories.  
Furthermore they are sensible locally; the near horizon behavior
($r \rightarrow 0$ limit) is the same as the usual D-particle solutions.    
The more detailed explanation of the relationship between these
two solutions is the theme of this subsection.  In what
follows, we will find that the satisfactory answers to the 
two issues raised in the above can naturally be given in 
the context of the type IIA/M-theory.  A clear hint
is provided by the inspection of the form of the functions
$f$ and $h_0$.  In the large charge and fixed distance limit,
the form of the function $f$ asymptotically becomes 
that of the function $h_0$.
Therefore, in view of the fact that the D-particle number $Q$ 
is the quantized Kaluzu-Klein momentum 
along the M-theory circle, the natural expectation is that
the solution (\ref{non-flat}) is the infinitely boosted
version of the solution (\ref{d0}) along the M-theory 
circle.  To see this more explicitly, we lift the
the D-particle solution (\ref{d0}) into the 
eleven-dimensional supergravity and rewrite
it as
\begin{equation}
ds^2_{11} = e^{-2 \phi /3} ds_{10}^2 +
            e^{4 \phi /3 } (dx_{11} - A_t dt )^2 
\label{eled0}
\end{equation}
\[ = - \frac{1}{f} dt^2 + f (dx_{11} - (1 - \frac{1}{f} )
      dt )^2 + dx_1^2 + \  \cdots \  + dx_9^2 \ , \]
where we use the standard dual description between the M-theory
on a circle and the IIA supergravity.  In our convention, the
M-theory circle, parameterized by $x_{11}$, has the period
of $R_s$ and we identify
\[ x_{11} \simeq x_{11} + R_s \ \ , \ \ t \simeq t \]
to produce the two dimensional cylinder parameterized
by $(x_{11} , t)$. \footnote{We closely follow the notation
and the idea of Ref. \cite{seiberg} in our paper with a slight
modification of the boost parameter due to our coordinate choice.}  
For our subsequent consideration, it is convenient to
introduce a set of asymptotic
light-cone coordinates $x^{\pm} = x_{11} \pm t$
and rewrite the metric Eq.(\ref{eled0}) in terms of $x^{\pm}$
as
\begin{equation}
ds_{11}^2 =  d x^+ dx^- + h dx^- dx^-
            + dx_1^2 + \  \cdots \  + dx_9^2
\label{temp1}
\end{equation}
where $h = f -1$.

Under a Lorentz boost along the M-theory circle
given by the boost parameter
\[ \beta = \frac{R}{\sqrt{R^2 + 4 R_s^2}} , \]
the original spatial circle with the period $R_s$ approaches the
light-cone circle with the period $R$, as we take the
limit $\beta \rightarrow 1$ and $R_s \rightarrow 0$ 
while keeping $R$ fixed.  To see this, we note that the
unboosted identification under the lattice translation
\[ x^+ \simeq x^+ + R_s \ \   , \ \  x^- \simeq x^- + R_s \]
changes into 
\[ x^+ \simeq x^+ + \frac{2}{1+ \sqrt{ 1 + 4 R_s^2 / R^2 }}
                    \frac{R_s^2}{R} \ \ , \ \ 
   x^- \simeq x^- + \frac{1+ \sqrt{ 1 + 4 R_s^2 / R^2 }}{2} R \]
under the boost.  We also find that the effect of the Lorentz 
boost on the metric corresponds to the substitution 
\begin{equation}
 h \ \rightarrow \  
 h_0 = \frac{4}{( 1 + \sqrt{1 + 4 R_s^2 / R^2} )^2 } 
    \frac{R_s^2}{R^2} \ h
\label{temp2}
\end{equation}
in Eq. (\ref{temp1}).  Under the infinite-boost,
the spatial circle becomes a light-cone circle
with the radius $R$ and we have a discrete light cone
where we identify $x^- \simeq x^- + R$. \cite{susskind}
Thus, to obtain 
a ten-dimensional, i.e., the type IIA side form of the 
infinitely boosted solutions,
we have to compactify Eq.(\ref{eled0}) along the 
asymptotic light-cone
circle.    Upon the dimensional reduction along the
light-cone circle $x^-$ with the radius $R$, the equations 
(\ref{temp1}) with $h$ given in Eq. (\ref{temp2}) reduces to 
Eq.(\ref{non-flat}) with 
\begin{equation} 
k =  Q / R^2  .
\end{equation}
The ten dimensional time coordinate $t$ now should be identified
with the light-cone time $t = x^+ /2$ as is clear from the
construction.
As expected, this consideration shows that the non-asymptotically 
flat solution (\ref{non-flat}) is the infinitely-boosted version 
of the D-particle solution along the M-theory circle.

The original D-particle solutions have the light-cone
energy $p_+ = 0$, due to the zero transversal momentum
and the zero rest mass of the graviton, and the momentum 
$p_- = Q/ R_s$.  After the infinite boost, the momentum
$p_-$ transforms into $p_- = Q / R$.  This can be
mostly clearly seen if we write the eleven dimensional
form of the (infinitely-boosted) solution Eq.(\ref{non-flat})
as
\begin{equation}
ds_{11}^2 =  dx^+ dx^- + h_0 dx_-^2 +  
              dx_1^2 + \  \cdots \  + dx_9^2   
\label{non1}
\end{equation}
\begin{equation}
             \simeq  dx^+ dx^- + \frac{p_-}{r^7}  \delta (x^- ) 
             dx_-^2  + dx_1^2 + \  \cdots \  + dx_9^2  .
\label{non2}
\end{equation}
Here $p_- = Q/R $ represents the momentum of the graviton
moving along the asymptotic light-cone circle, 
$h_0 = \ Q/ (R^2 r^7) $, and we replaced $1/R$
factor with the delta function defined on the finite range
$0 \le x^- \le R$.  Eq. (\ref{non2}) is the eleven dimensional
Aichelberg-Sexl metric \cite{aich} representing 
the gravitational shock-wave 
travelling along the light-cone circle with the momentum $p_-$, 
as was pointed out in Ref. \cite{becker}.  In the Kaluza-Klein
dimensional reduction from eleven dimensions to ten dimensions,
zero mode part of the solution Eq.(\ref{non1}) is the same
as that of Eq.(\ref{non2}).  The difference comes only from the 
massive higher modes, justifying our approximate identification
of these two equations.  

Our consideration so far provides us with the answers to the 
questions we posed earlier.  First, when we take the limit 
$k \rightarrow 0$, the eleven dimensional version Eq.(\ref{non1})
of the solution Eq.(\ref{non-flat}) indeed
becomes a eleven dimensional flat solution, even if 
Eq.(\ref{non-flat}) has a singular limit.  We note that we
write Eq.(\ref{non1}) as
\[ ds_{11}^2 = h_0^{-1 /2} ( - \frac{1}{ \sqrt{h_0}} d (
  \frac{x^+}{2} )^2 
  + \sqrt{h_0} ( dx_1^2 + \  \cdots \  + dx_9^2 ) )
  + h_0 ( dx^- + \frac{1}{h_0} d ( \frac{x^+}{2} ) )^2 . \]
for the dimensional reduction along the light-cone circle.
Thus, the singular limits of the ten dimensional
dilaton, graviton and
graviphoton fields combine to give a well-behaved
eleven dimensional limit, as we take $k \rightarrow 0$
(thereby $h_0 \rightarrow 0$).  Just as the $Q=0$ D-particle 
solution gives the eleven dimensional flat space-time
with a compact spatial circle, $k=0$ non-asymptotically
flat solution Eq.(\ref{non-flat}) gives the eleven
dimensional flat space-time with a compact light-cone circle.
Secondly, we now have a better understanding of the energy
of the non-asymptotically flat solutions Eq.(\ref{non-flat}).
In the case of the D-particles, the energy computed from the
eleven dimensional perspective is just the energy 
$E= Q/ R_s$ of the graviton travelling along the spatial circle,
which is conjugate to the time coordinate $t$ in 
Eq.(\ref{eled0}).  This time coordinate $t$ is identical to
the ten dimensional time and we thus recover the energy of
D-particles, $E = Q /(gl_s )$.  In the similar spirit, since
the solutions Eq.(\ref{non1}) are also asymptotically flat
from the eleven dimensional point of view,
we can unambiguously compute the light-cone energy $p_+ = 0$, 
both before and after the boost.      
Since the ten dimensional time in the case of non-asymptotically
flat solutions corresponds to $x^+ /2$, its conjugate energy is
again $2p_+ = 0$.  This argument shows that the assignment of
zero energy to the non-asymptotically flat solutions 
Eq.(\ref{non-flat}) in ten
dimensions is a natural one.

\section{Type IIB supergravity on five-torus with 
intersecting D-Strings and D 5-branes}

Following the procedure explained in Section II, we can turn
an asymptotically flat solution to a non-asymptotically
flat one via the infinite boost along the M-theory circle.
Using a chain of $T$-dualities, we can obtain corresponding
non-asymptotically flat solutions for any D $p$-brane solutions,
given our analysis of the D-particle solutions.
The case of interest in this section, in particular, is the
non-asymptotically flat solutions from the (non)-extremal
solutions of 
the type IIB supergravity on a five-torus with intersecting
D-strings and D 5-branes and with the Kaluza-Klein momentum
along the circle where D-strings are wrapped.  These
configurations of D-branes were used in the computation
of the black hole entropy via the D-brane technology,
and they provide one of the simplest setting to discuss
the quantum black hole physics. \cite{cm}  By 
performing $U$-dual
transformations to these well-understood configurations,
we can better understand other configurations of
D-branes and NS-branes.  The question is then the
action of the infinite boost along the M-theory circle
on these configurations.

We consider D-strings wrapped along the circle
parameterized by $x_9$, D 5-branes wrapped 
on the five-torus parameterized by $x_5$,
$\cdots$, $x_9$, and the Kaluza-Klein momentum
moving along the circle $x_9$.   The charges produced
by D-branes in the non-compact five dimensions ($t$, $x_1$,
$\cdots$, $x_4$) are related to two real numbers
$r_1$ and $r_5$, respectively.   
The metric, dilaton, and R-R two-form gauge field
$A_{\mu \nu}$ are given by \cite{cvetic}
\begin{equation}
ds_{10}^2 = \frac{1}{\sqrt{f_1 f_5}} ( -dt^2 + dx_9^2 
+ K (\cosh \sigma dt - \sinh \sigma dx_9) ^2 ) 
\label{d15}
\end{equation}
\[ + \sqrt{f_1 f_5}
( \frac{1}{1 - K} dr^2 + r^2 d \Omega_3^2 )
+ \sqrt{\frac{f_1}{f_5}} (dx_5^2 + \
\cdots \ + dx_8^2 ) ,  \]
\[ e^{-2 \phi} = \frac{f_5}{f_1}  \ \ , \ \
    A_{t x_9} = 1  - \frac{1}{f_1} \ \ , \ \ 
    (dA)_{ijk} = \frac{1}{2} \epsilon_{ijkl} \partial_l f_5 \]
where we introduce three functions
\[ f_1 = 1 + \frac{r_1^2}{r^2} \ \ , \ \ 
      f_5 =  1 + \frac{r_5^2}{r^2} \ \ , \ \ 
      K = \frac{r_0^2}{r^2}  .\]
Here greek indices and latin indices are ten-dimensional
and non-compact spatial four-dimensional, respectively.  
For non-compact spatial dimensions, we use the radial coordinate $r$
which becomes $r^2 = x_1^2 + x_2^2 + x_3^2 + x_4^2$ in the
extremal limit, and 
$d \Omega_3^2$ denotes the unit three-sphere.  For the extremal
solutions, the real numbers $\sigma$ and $r_0$ take the
limit $\sigma \rightarrow \infty$ and $\ r_0 \rightarrow
0$ while keeping $r_0^2 \sinh 2 \sigma $ fixed.  

We now consider the change of the
function $f_5$ of D 5-branes under the analysis of the
section II.  For this purpose, we first apply T-dual transformations
to all of the compact coordinates from $x_5$ to $x_9$.
These transformations turn the D 5-branes into D-particles,
the D-strings into D 4-branes wrapped on a four torus
parameterized by $x_5$, $\cdots$, $x_8$, and the Kaluza-Klein
momentum into fundamental strings winding the $x_9$
circle.  Secondly, we lift the resulting ten dimensional
type IIA solution into the eleven dimensional solution.
>From the eleven dimensional perspective, we have
longitudinal 5-branes (originally D-strings), 
longitudinal membranes (originally Kaluza-Klein momentum),
and  the momentum along the M-theory circle
(originally D 5-branes).  Straightforward calculation
of the transformations from Eq. (\ref{d15})
shows that we have
\begin{equation}
ds_{11}^2 = \frac{1}{f_1^{1/3} (1 + K \sinh^2 \sigma)^{2/3} }
      \{ ( 1+ \frac{K}{2 f_5 } ) dx^+ dx^- + ( h
         + \frac{K}{4 f_5 } ) dx^- dx^- 
         + \frac{K}{4 f_5} dx^+ dx^+   \}
\label{temp11}
\end{equation}
\[  + \frac{f_1^{2/3}}{(1 + K \sinh^2 \sigma)^{2/3} } dx_9^2
  + f_1^{2/3} (1 + K \sinh^2 \sigma )^{1/3} 
     ( \frac{1}{1 - K} dr^2 + r^2 d \Omega_3^2   ) \]
\[   + \frac{(1+K \sinh^2 \sigma )^{1/3}}{f_1^{1/3}} 
     (dx_5^2 + \ \cdots \ dx_8^2 )   \]
for the eleven dimensional metric where we introduce
$h = f_5 -1 $, quite similar to Eq. (\ref{non1}).
Now we perform the infinite-boost along
the M-theory circle while taking $R_s \rightarrow 0 $.
As in section II, $h$ represents D-particles and thus $h$ is of the
order of $O(R_s^{-2})$.  Since $K$ is the contribution 
from the Kaluza-Klein momentun along the $x_9$ direction,
we can assume that the value of its charge is smaller than 
the order of $O(R_s^{-1} )$.   Under the infinite boost,
the expression in the curly bracket of Eq. (\ref{temp11})
transforms into
\begin{equation}
   ( 1+ \frac{K}{2 f_5 } ) dx^+ dx^- +  \frac{R_s^2}{R^2} 
         ( h + \frac{K}{4 f_5 } ) dx^- dx^- 
         + \frac{R^2}{R_s^2} \frac{K}{4 f_5} dx^+ dx^+  
\label{temp12}
\end{equation}
\[ \simeq  dx^+ dx^- + h_5 dx^- dx^- 
      + \frac{K}{4 h_5} dx^+ dx^+    .\]
In obtaining the second line from the first line in the above
equation, we retain only the
leading order term in $R_s$, which turns out to be the
zeroth order in $R_s$.  We note that 
$h_5 =  ( R_s^2 / R^2 ) h $ is the zeroth order in $R_s$. 
Since the M-branes in our consideration are all longitudinal,
the functions $f_1$ and $K$ do not change.   
The additional effect of this boost is to change the spatial 
circle into the light-cone circle along the $x^-$ direction.
Thus, we plug 
Eq. (\ref{temp12}) into Eq. (\ref{temp11}) and compactify
the boosted eleven dimensional solution along the light-cone circle
to get the corresponding type IIA solution.  As a final step,
after the T-dual
transformations to each of the circles of the five-torus,
we go back to the original type IIB theory.  The fields
now look
\begin{equation}
ds_{10}^2 = \frac{1}{\sqrt{f_1 h_5}} ( -dt^2 + dx_9^2 
+ K (\cosh \sigma dt - \sinh \sigma dx_9) ^2 ) 
\label{d15p}
\end{equation}
\[ + \sqrt{f_1 h_5}
( \frac{1}{1 - K} dr^2 + r^2 d \Omega_3^2 )
+ \sqrt{\frac{f_1}{h_5}} (dx_5^2 + \
\cdots \ + dx_8^2 ) ,  \]
\[ e^{-2 \phi} = \frac{h_5}{f_1}  \ \ , \ \
    A_{t x_9} = 1  - \frac{1}{f_1} \ \ , \ \ 
    (dA)_{ijk} = \frac{1}{2} \epsilon_{ijkl} \partial_l h_5 \]
where we use
\[ h_5 = \frac{R_s^2}{ R^2 } (f_5 -1 ) = 
     \frac{\bar{r}_5^2}{r^2}. \]
Eq. (\ref{d15p}) is identical to Eq.(\ref{d15}), except
for the fact that the function $f_5$ is changed into 
the function $h_5$.  For a generic non-extremal solutions,
for which we can not use the BPS equations, it is not immediately
apparent that Eq. (\ref{d15p}) satisfies the field equations
of the type IIB supergravity.  However, we can verify this fact
by straightforward calculations. \footnote{As a verification 
of this, we can resort to Ref. \cite{kpl} where the 
general $s$-wave sector static
solutions of the type IIB supergravity on $T^5$ 
for the same brane and internal 
momentum configurations are given.  In fact, by setting
$\sinh (\sqrt{c_1} (\tilde{c} - \tilde{c}_1 )) = 0$
for the solutions reported in that paper, we can indeed confirm
that the ten dimensional field equations are satisfied for
Eq. (\ref{d15p}).}    Since $h_5$ approaches zero 
as $r$ goes infinity, the solution Eq. (\ref{d15p}) is clearly
not asymptotically flat.  

Further properties of the non-asymptotically flat solutions
can be learned if we take the $S$-dual transformation
of the type IIB theory solution.  The original D-strings 
and D 5-branes
described by Eq.(\ref{d15}) become the fundamental 
strings and NS 5-branes, respectively, under the $S$-dual
transformations.  In the case of the solution Eq. (\ref{d15p}),
it transforms into
\begin{equation}
ds_{10}^2 =  \frac{1}{f_1} ( -dt^2 + dx_9^2 
+ K (\cosh \sigma dt - \sinh \sigma dx_9) ^2 ) 
 + \frac{\bar{r_5^2}}{r^2 ( 1-K )} dr^2  
\label{smet}
\end{equation}
\[  + \bar{r}_5^2 d \Omega_3^2 + (dx_5^2 + \
    \cdots \ + dx_8^2 ) , \]
\[ e^{-2 \phi} = \frac{f_1}{h_5}  \ \ , \ \
    B_{t x_9} = 1  - \frac{1}{f_1} \ \ , \ \ 
    (dB)_{ijk} = \frac{1}{2} \epsilon_{ijkl} \partial_l h_5 \]
under the $S$-dual transformation, where $B_{\mu \nu}$
is the NS-NS two-form gauge field.  The solution 
Eq. (\ref{smet}) is again non-asymptotically flat
in the five dimensional space-time.  In fact,
we observe that a spontaneous compactification of 
the would-be non-compact coordinates occurs.
The metric Eq. (\ref{smet}) shows that the ten dimensional 
manifold ${\cal M}_{10}$ becomes a tensor product
${\cal M}_{10} = {\cal M}_{3} \times S^3 \times T^4$,
where $S^3$ is a three-sphere ($d \Omega_3^2$) with a 
fixed radius and $T^4$ is a four-torus parameterized by
$x_5$, $\cdots$, $x_8$ with fixed torus moduli.

We can further compactify the metric Eq.(\ref{smet})
upon the $x_9$ circle to get the two dimensional 
solutions.  These solutions turn out to be the two-dimensional
black holes studied in Ref. \cite{mny}.  Furthermore, if we set
$r_1 = \sigma = 0$, we end up exactly recovering the CGHS black
hole, where $\bar{r}_5^2$ plays the role of the
cosmological constant. \cite{cghs}  We can easily see this by comparing 
Eq. (\ref{smet}) with the metric of the CGHS model in a coordinate
system where the radial coordinate $r$ is related to the dilaton $\phi$ 
via $r = \exp ( - \phi )$.  When we approach the extremality
with the vanishing Kaluza-Klein momentum and the vanishing $r_1$, we 
have the linear dilaton vacuum, which is the vacuum of the 
CGHS model and clearly has zero ADM mass.  
This behavior is consistent with our previous mass assignment; 
the mass contribution from $\bar{r}_5^2$, after 
the infinite boost along the M-theory circle, vanishes.
On the contrary, for the NS five-branes,
the same conditions $r_1 =0$ and the vanishing Kaluza-Klein
momentum at the extremality give 
the ADM mass proportional to $r_5^2$ according to the 
BPS mass formula.
 
\section{Concluding Remarks}

One interesting result from our analysis is that certain black
holes in differing space-time dimensions are connected to each other
via $U$-dualities and the eleven dimensional diffeomorphism
(the infinite Lorentz boost along the M-theory circle being
part of it).  This aspect is consistent with the idea of 
\cite{univ} where it is shown that the essential features of the low
energy effective string theories are universal.  If this universality 
and the relationship between higher and lower dimensional
black holes are confirmed at the level of
the dynamics, we will be able to have better understanding
of this universal behavior from the relatively easier
investigation of the lower dimensional gravity theories
(such as the CGHS model) tensored with appropriate
internal theories.  In this process, some non-asymptotically
flat solutions we discussed in this paper are expected
to play a role, and we have shown that some bothering 
features of them can be understood in the framework
of the type IIA/M theory.  Given our results, one urgent
next step will be to study the (quantum) dynamics 
happening on the background of the non-asymptotically
flat solutions.  Related to this issue is the recent work
\cite{malda}, where the solution Eq. (\ref{smet}) 
with $r_1 = 0$ and $\sigma= 0$ describes the near horizon 
dynamics of the transversal five-branes and it is shown that
the semiclassical analysis is reliable in some regimes
of the parameters.

\acknowledgements{Y.K. would like to thank Dahl Park
for useful discussions. S.H. is supported in part by Korea Research
Foundation.}

\end{document}